\documentclass[onecolumn]{elsart}
\usepackage{graphicx}
\usepackage{dcolumn}
\usepackage{bm}
\usepackage{amsmath}
\input epsf
\epsfclipon

\begin{document}
\begin{frontmatter}
\title{Statistical properties of short term price trends in high frequency stock market data}
\author{Pawe\l \ Sieczka} and
\author{Janusz A. Ho{\l}yst\corauthref{holyst}}
\corauth[holyst]{Corresponding author. Tel.: +48 22 234 71 33; fax: +48 22 234 58 08}
\ead{jholyst@if.pw.edu.pl}
\address{Faculty of Physics and Center of Excellence
for Complex Systems Research, Warsaw University of Technology,
Koszykowa 75, PL-00-662 Warsaw, Poland}
\date{\today}

\begin{abstract}
We investigated distributions of short term price trends for high frequency stock market data. A number of trends as a function of their lengths was measured. We found that such a distribution does not fit to results following from an uncorrelated stochastic process. We proposed a simple model with a memory that gives a qualitative agreement with real data.
\end{abstract}

\maketitle
\begin{keyword}
Econophysics \sep Financial markets \sep Price trends
\PACS 89.65.Gh
\end{keyword}
\end{frontmatter}

\section{Introduction}
Statistical analysis of stock prices is a rich source of information about the nature of financial markets. It was Louis Bachelier who used a stochastic approach to model financial time series for the first time \cite{Bachelier}. Since that time the statistical analysis of stock prices has become a widely investigated area of interdisciplinary researches \cite{Mantegna0,Bouchaud,APFA4,EPJB}. 

In 1973, Fischer Black and Myron Scholes published their famous work \cite{Black}  where they presented a model for pricing European options.  They assumed that a price of an asset can be described by a geometric Brownian motion. However, the behaviour of real markets differs from the Brownian property \cite{Mandelbrot,Fama1}, since the price returns form a truncated L\'evy distribution \cite{Mantegna1,Mantegna2,Kertesz}. As a result of this observation many non-Gaussian models were introduced \cite{Mantegna0,Bouchaud,APFA4}.

Another divergence from Gaussian behaviour is an autocorrelation in financial systems.  Empirical studies show that the autocorrelation function of the stock market time series decays exponentially with a characteristic time of a few minutes, while the autocorrelation of prices absolute values decays slower, as a power law function, what leads to a volatility clustering \cite{Stanley,Stanley2,Stanley3,Krawiecki}. 

The issue of market memory was also considered by many authors (see references in \cite{Mantegna0,Bouchaud}). It was observed \cite{Fama2}, that for certain time scales, a sequence of two positive price changes leads more frequently to a subsequent positive change than a sequence of mixed changes, i.e. the conditional probability $P(+\backslash++)$ is larger than $P(+\backslash+-)$.  In this paper we investigated this effect for high frequency stock market data.

\section{Empirical data}
Let us consider short term price trends for  high frequency stock market data. 
By {\it short term uptrend/downtrend} we mean such a sequence of prices that a price is larger/smaller than the preceding one (see below for a more precise definition).

First, having a time  series $Y_t$, which is in our case a history of a stock price or a market index, we build a series of variables $s_t$ in the following way:
\begin{itemize}
\item $s_t=1$ if $Y_t>Y_{t-1}$,
\item $s_t=-1$ if $Y_t<Y_{t-1}$,
\item $s_t=s_{t-1}$ if $Y_t=Y_{t-1}$.
\end{itemize}
A positive value of the variable $s_t$ means that at the time $t$ the price $Y_t$ did not decrease, and similarly a negative value means that the price did not increase.

In  a series $s_t$ we can distinguish subseries of identical values. For $a<b$ and $s_a=s_b=s$, $S(a,b,s)$ is such a subseries if and only if $\forall_{c\in (a, b)} \, s_c=s$. 
Subseries $S(a,b,s)$ can be identified with an uptrend lasting from $t=a$ till $t=b$ for $s=1$, and with a downtrend for $s=-1$. The length $l$ of such a uptrend/downtrend is equal to $b-a+1$. Let us mention that a subseries of a length $l$ includes two subseries of length $l-1$, three subseries of length $l-2$ etc.  

Let $N(l)$ be a number of subseries of a length $l$ with a fixed $s$ in a series $s_1, ..., s_M$. If $s_t$ were generated by an uncorrelated discrete stochastic process with a probability $P(s_t=1)=p$, then the expected value of $N(l)$ would be equal to:
\begin{equation}\label{Wiener}
N(l)=(M-l+1)p^l,
\end{equation}
where $M$ is a number of all elements in the basic series. Similarly the expected value of downtrend series of length $l$ is $N(l)=(M-l+1)(1-p)^l$.

\begin{figure}
\includegraphics[scale=0.5]{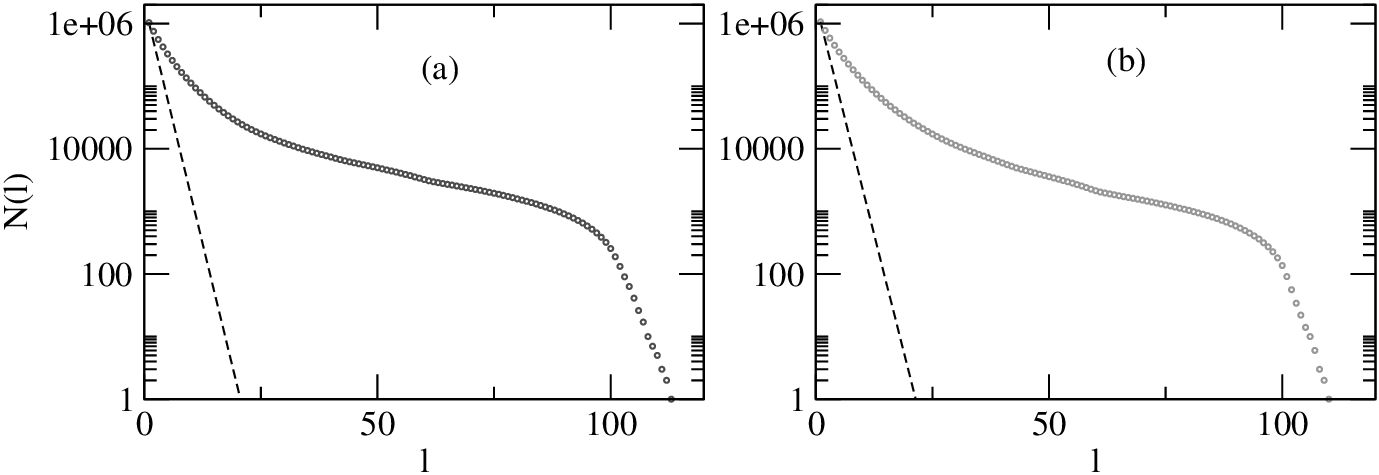}
\caption{Distribution $N(l)$ for uptrends (a) and downtrends (b) for the WIG20 index between  the 13th June 2003 and the 3rd November 2006. Data are sampled every 15 seconds (circles), and were compared to uncorrelated process (line) eq. (\ref{Wiener}).} 
\label{fg1}
\vspace{1cm}
\end{figure}

\begin{figure}
\includegraphics[scale=0.5]{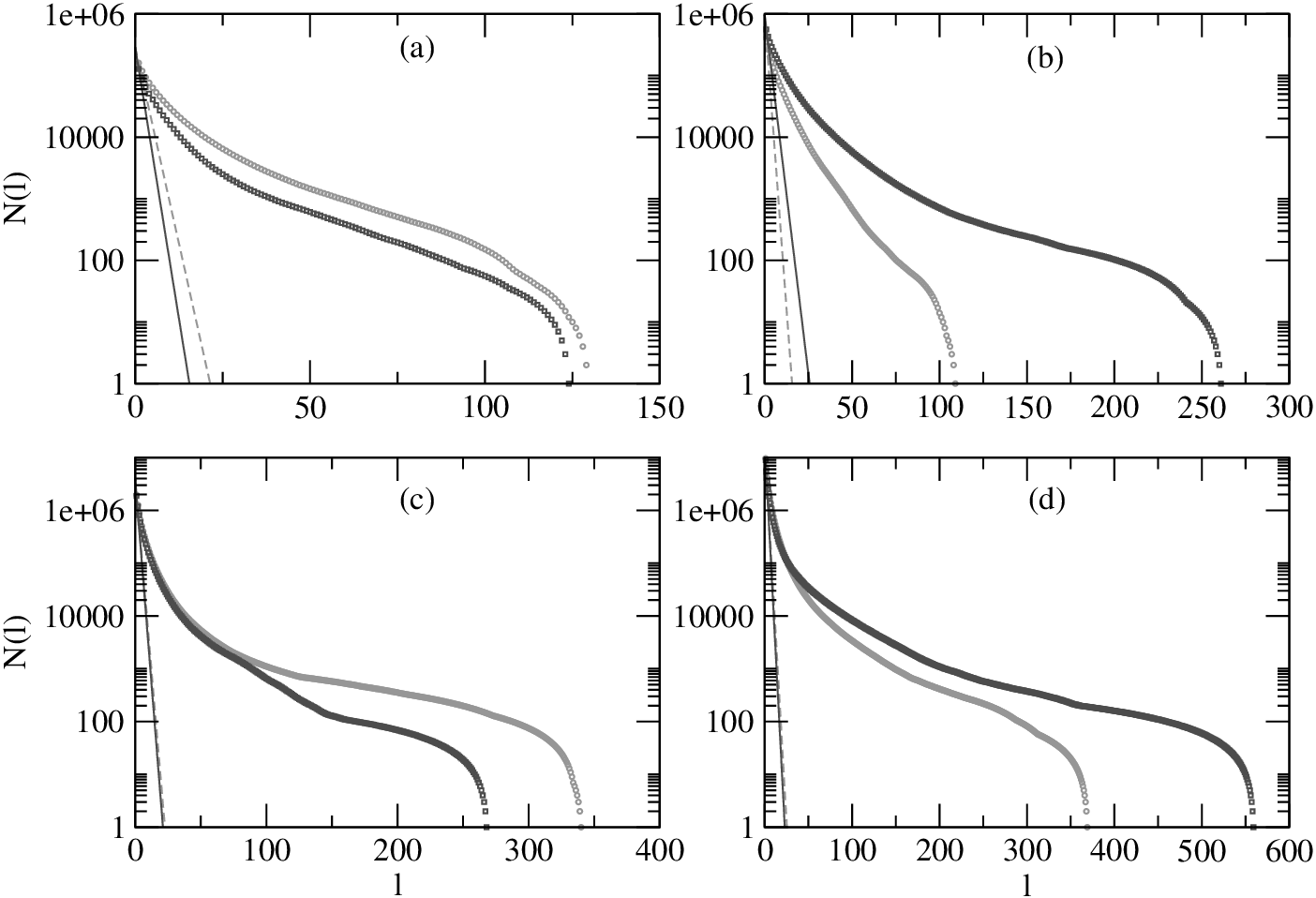}
\caption{Distribution $N(l)$ for: (a) BIOTON between the 31st March 2005 and the 3rd November 2006 (WSE), (b) TPSA between the 17th November 2000 and the 3rd November 2006  (WSE), (c) APPLE between the 4th January 1999 and the 29th December 2000 (NYSE), (d) INTEL between the 4th January 1999 and the 29th December 2000 (NASDAQ). Uptrends are plotted with circles and downtrends are plotted with squares. All data are sampled tick by tick. Lines correspond to the uncorrelated process (\ref{Wiener}).}
\label{fg2}
\vspace{0.5cm}
\end{figure}

We have measured the distribution $N(l)$ for real market data and the same distribution for the corresponding uncorrelated process. Figure \ref{fg1} pre\-sents this distribution for the WIG20 index of Warsaw Stock Exchange (WSE) between the 13th June 2003 and the 3rd November 2006, and the distribution for the corresponding uncorrelated process. The results show a significant difference between real data and the uncorrelated model. If variables $s_t$ were uncorrelated, there would not be subseries longer than 25 ticks. In fact, subseries even longer than 100 ticks are present. The trends last for about 30 minutes. There are far more such trends than it would be if the process were uncorrelated. The distribution $N(l)$ was also calculated for particular stocks from WSE, NYSE and NASDAQ (fig. \ref{fg2}). The stocks from WSE were: Bioton between the 31st March 2005 and the 3rd November 2006, and TPSA between the 17th November 2000 and the 3rd November 2006. The stock from NYSE was Apple, and the stock from NASDAQ was Intel, both between the 4th January 1999 and the 29th December 2000.  For the index WIG20 trend periods were measured in real time, but for the stocks they were measured in a transaction time (see section 4).

\begin{figure}
\vspace{0.5cm}
\includegraphics[scale=0.45]{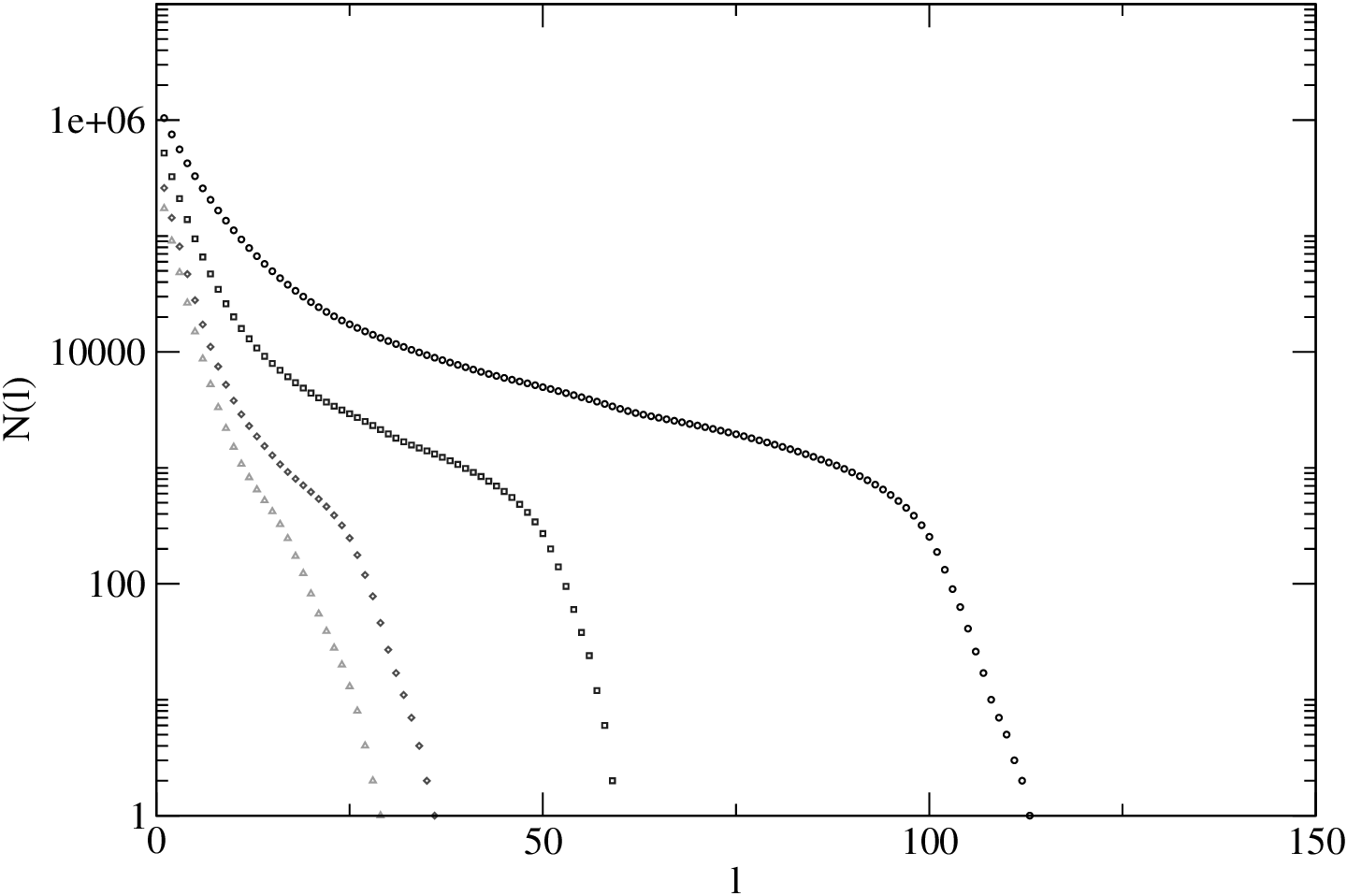}
\caption{$N(l)$ of WIG20 uptrends for every n-th element of $s_t$, n=1 circles, n=2 squares, n=4 diamonds, n=6 triangles.\vspace{1cm}}
\label{fg3}
\end{figure}

The observed difference between the uncorrelated model and the real markets is due to strong autocorrelations in the process $s_t$. It is only seen in high frequency data. Choosing every n-th element of the series $s_t$ weakens the autocorrelations, and makes the outcome approaching the uncorrelated model with growing n. It is shown in fig. \ref{fg3}.

\section{A phenomenological model of correlated market prices}
In real markets variables $s_t$ and $s_{t+\tau}$ are correlated, although this correlations decay very fast. Let $r(k)$ stand for a conditional probability $P(s_{n+k+1}=1 \backslash s_{n+k}=1, ..., s_n=1, s_{n-1}=-1 )$, which is independent of $n$. For processes where autocorrelations are present we can write a generalization of equation (\ref{Wiener}):
\begin{equation}\label{Nl}
N(l+1)=(M-l)p\prod_{i=1}^l r(i),
\end{equation}
for $l>0$ and $N(1)=Mp$.

Let us see that the result (\ref{Nl}) is equivalent to (\ref{Wiener}) if for any $x$ there is $r(x)=p$. A key issue is to model $r(x)$ in order to describe  characteristics of a given market.  The results presented in fig. \ref{fg1}-\ref{fg2} show that the model of the uncorrelated process (\ref{Wiener}), is a poor simplification. To get a better consistency with real data the function $r(x)$ can be modelled as:
\begin{equation}\label{rx}
r(x)=a(x-x_1)(x-x_2),
\end{equation}
with fitted parameters $a$, $x_1$ and $x_2$. A binomial function was chosen because we are looking for a simple concave function with a maximum, and a binomial function matches our requirements for proper parameters $a$, $x_1$, $x_2$. 

We expect that for small $x$, the probability value $r(x)$ increases with $x$. It means that when the trend starts forming, investors follow it and, as a result, they amplify the trend. Thus, the probability of a continuation of the price movement  grows.   As time goes by, some of them may want to withdraw to take profits, and those who are out of the market believe it is to late to get in. This causes  a decrease of $r(x)$ for a longer trend.

\begin{figure}
\includegraphics[scale=0.55]{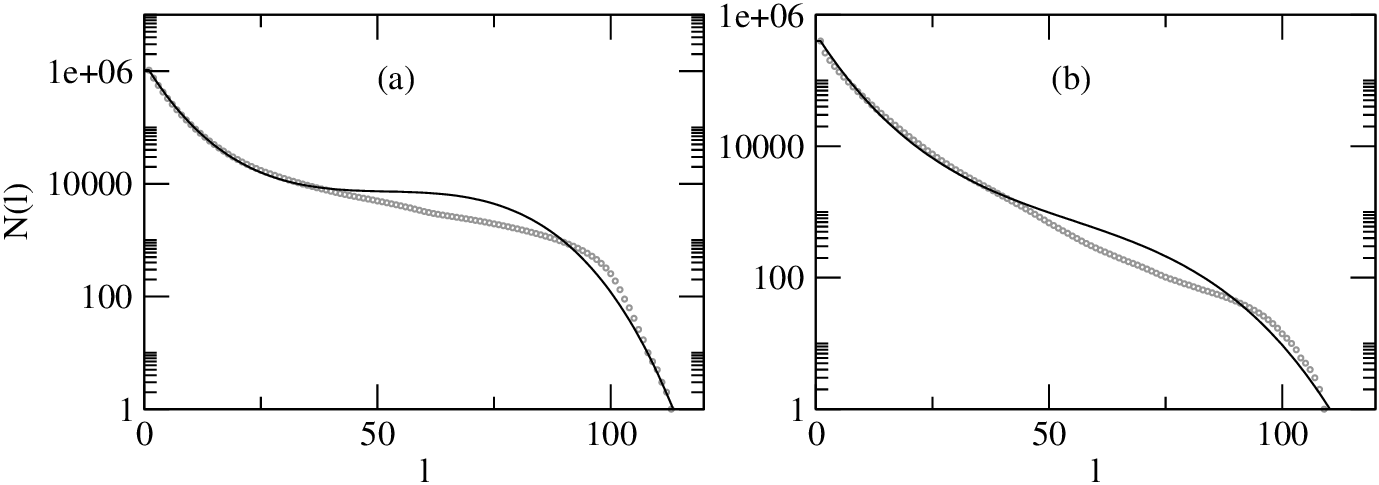} 
\caption{The distribution $N(l)$ with fitted curve (\ref{Nl}) (a) WIG20 (WSE) uptrends with fitted parameters: $a=-0.000098$, $x_1=-48.28$, $x_2=153.31$, (b) TP S.A. company (WSE) uptrends with fitted parameters: $a=-0.000057$, $x_1=-73.42$, $x_2=184.35$.}
\label{fg4}
\end{figure}

One can choose various functions to model the probability $r(x)$. All such functions should be concave functions with a maximum for a positive argument smaller than the maximal length of all subseries.
The figure \ref{fg4} presents the  distribution  $N(l)$ with a fitted curve based on (\ref{rx}).

\begin{figure}
\includegraphics[scale=0.45]{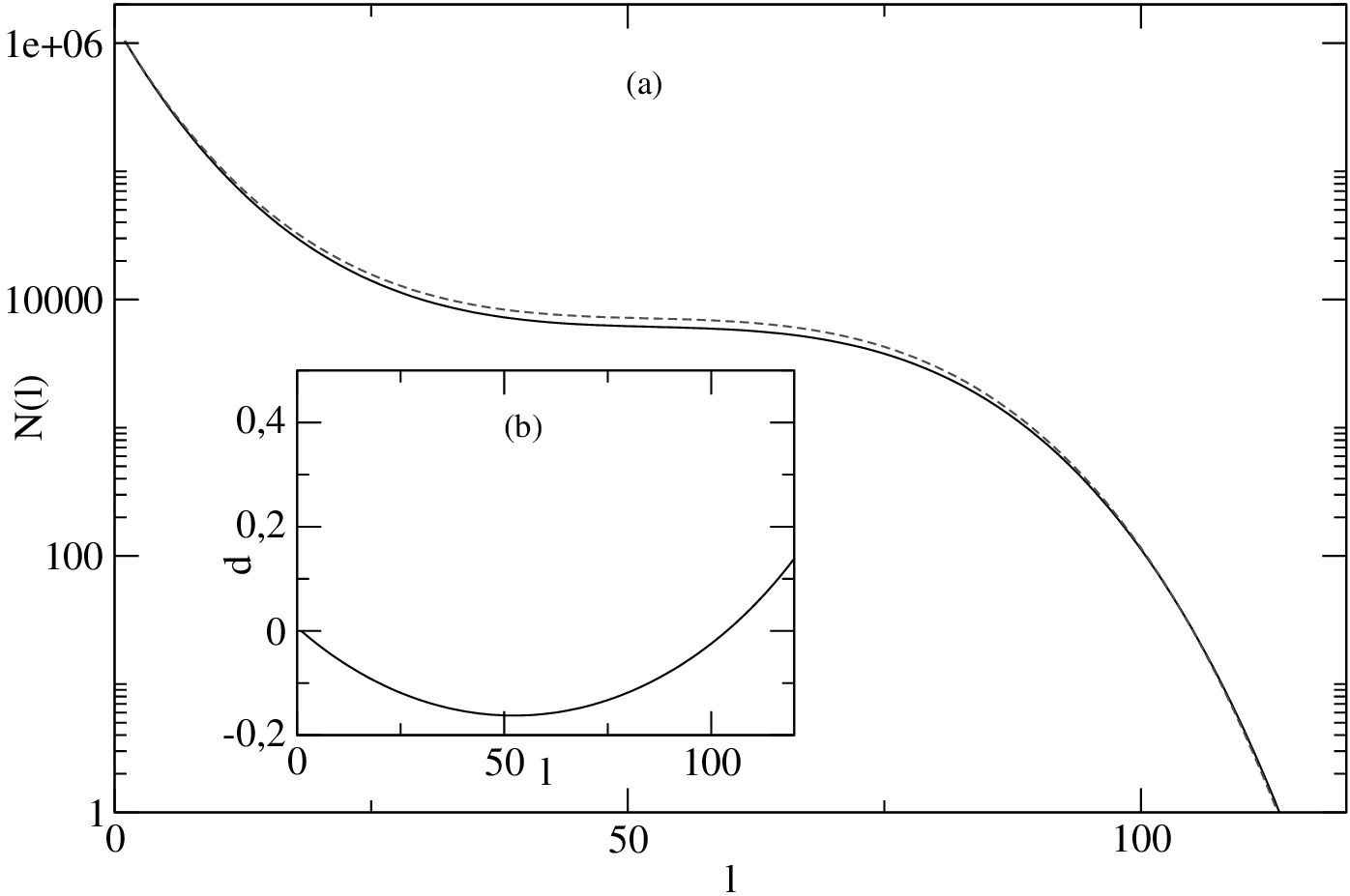} 
\caption{Graph (a) presents the distribution $N_1(l)$ obtained from the equation (\ref{Nl}) and $N_2(l)$ from the approximation (\ref{Nl-approx}) for the WIG20 index (WSE). Graph (b) presents the relative difference between them: $d=(N_2(l)-N_1(l))/N_2(l)$.\vspace{1cm}}
\label{wykres2}
\end{figure}

Putting (\ref{rx}) into (\ref{Nl}) we get after some algebra an approximated form of the  function $N(l)$ as:
\begin{equation}\label{Nl-approx}
\begin{split}
N(l)\simeq(&M-l+1)p{\rm e}^{-2(l-1)}[a(l-x_1)(l-x_2)]^{l-1}\\ \times & \Big (\frac{l-x_1}{1-x_1}\Big )^{1-x_1}\Big (\frac{l-x_2}{1-x_2}\Big )^{1-x_2}.
\end{split}
\end{equation}
The figure \ref{wykres2} presents functions (\ref{Nl}) and (\ref{Nl-approx}) and a relative difference between them.

\begin{figure}
\vspace{0.5cm}
\includegraphics[scale=0.45]{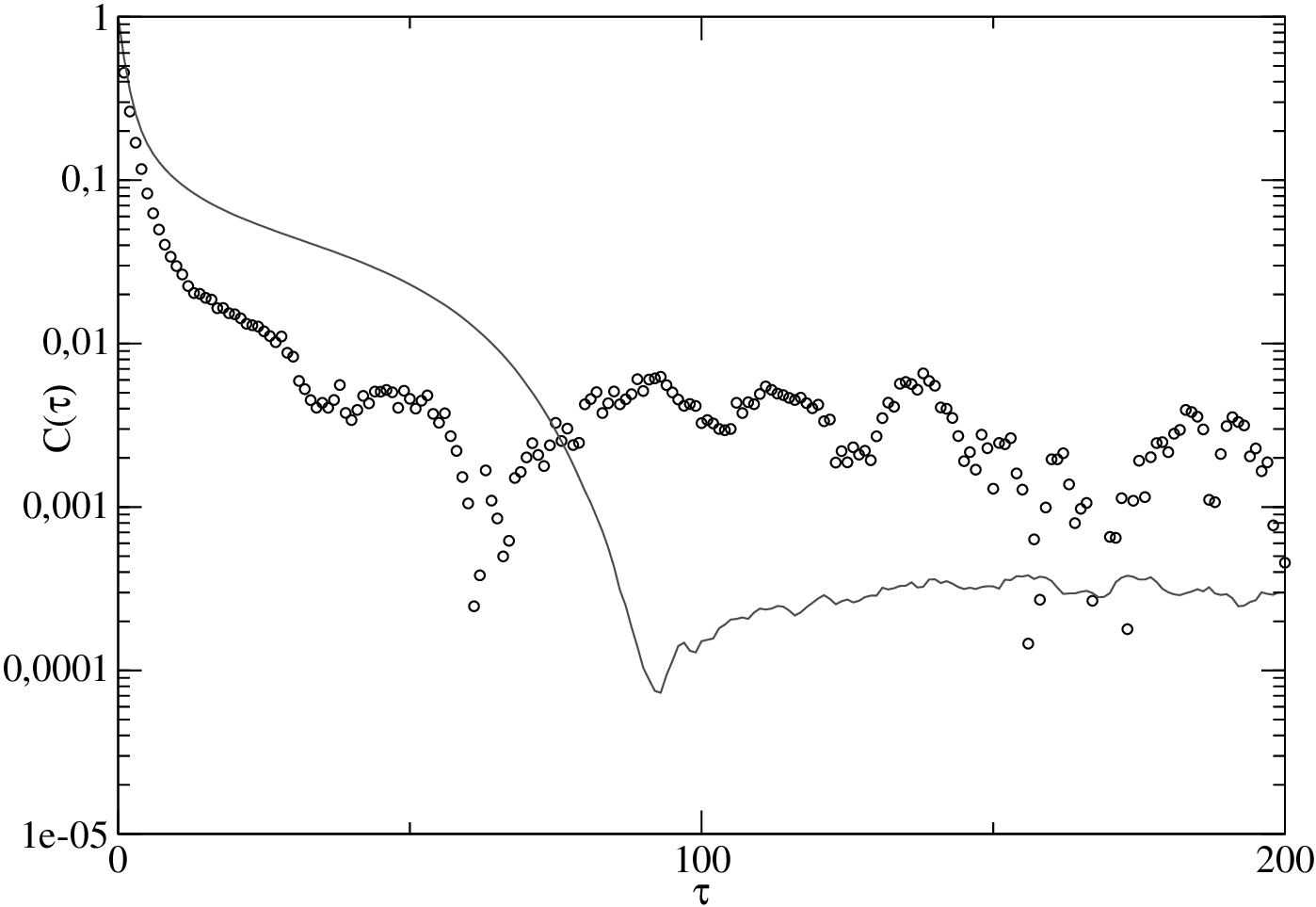} 
\caption{The autocorrelation function obtained from eq. (\ref{corr})  for parameters $a^+=-0.000098$, $x_1^+=-48.28$, $x_2^+=153.31$, $a^-=-0.000101$, $x_1^-=-47.32$, $x_2^-=150.69$ (line) and the index WIG20 itself (circles).\vspace{1cm}}
\label{wykres3}
\end{figure}

For a given set of parameters $a^+=-0.000098$, $x_1^+=-48.28$, $x_2^+=153.31$, $a^-=-0.000101$, $x_1^-=-47.32$, $x_2^-=150.69$, obtained for uptrends and downtrends in the WIG20 index respectively, one can simulate the stochastic process according to eq. (\ref{Nl}, \ref{rx}). The autocorrelation function
\begin{equation}\label{corr}
C(\tau)=\langle s_{i+\tau} s_i\rangle,
\end{equation}
calculated for such a process, decrease similarly to the autocorrelation function received from empirical data (see fig. \ref{wykres3}). The model reflects short range correlations of the sign.

\section{Measuring trends in volume and volatility times}

In previous chapters we presented an analysis of price trends measured in real and transaction times.  The WIG20 index is published every 15 seconds, and for this index the 15-seconds data are the most frequent possible. Investigating these data we naturally used the real time with a 15-seconds interval, in which the sequence "$+1, +1, +1, +1$" means that the index did not decrease during one minute.

For the stocks of companies tick-by-tick data are accessible, thus  a transaction time is a natural measure of a time length. Transaction time can be defined as:
\begin{equation}
\tau_t(t_{i})=\tau_t(t_{i-1})+1,
\end{equation}
where $t_i$ is the real time of the transaction $i$. We used  times $\tau_t$ for data analysis of single stocks. In this case the sequence "$+1, +1, +1, +1$" means that the price did not decrease during four subsequent transactions.

Other time definitions are also possible. One of them is the volume time \cite{Farmer} defined in a standard way as:
\begin{equation}
\label{tau_v}
\tau_v(t_i)=\tau_v(t_{i-1})+V_i,
\end{equation}
where $V_i$ is the volume of transaction $i$. 

\begin{figure}
\vspace{0.5cm}
\includegraphics[scale=0.5]{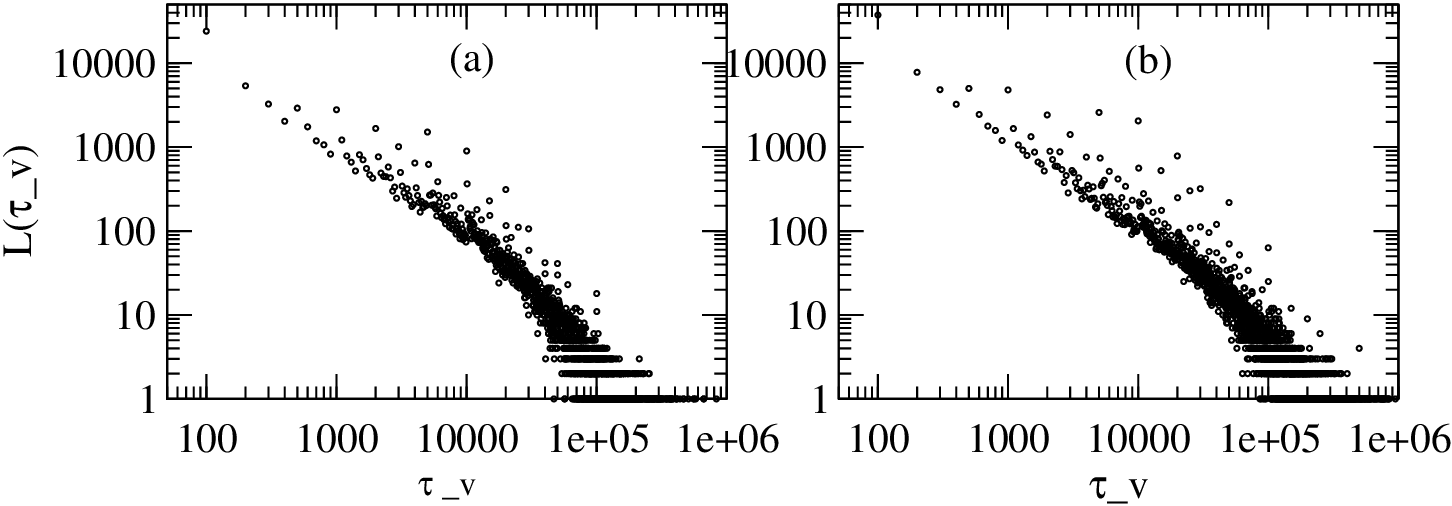} 
\caption{The distribution $L(\tau_v)$ of numbers of trend periods of a length $\tau_v$ measured in a volume time, for: (a) PKN Orlen (WSE), (b) TPSA (WSE). Data were binned with a bin of the width $100$. \vspace{1cm}}
\label{wykres4}
\end{figure}

We repeated our analysis, measuring lengths of trend periods, using the volume time instead of the transaction time. In figure \ref{wykres4} we presented distributions of the trend periods of exact length $\tau_v$ ($L(\tau_v)$) for stocks PKN Orlen and TPSA of Warsaw Stock Exchange, both from the period between the 17th November 2000 and the 3rd November 2006. Let us stress that contrary to the distribution $N(l)$ which was, in a sense, cumulative, $L(\tau_v)$ is non-cumulative, because it shows only trends of the {\it exact} length $\tau_v$. 

\begin{figure}
\vspace{0.5cm}
\includegraphics[scale=0.5]{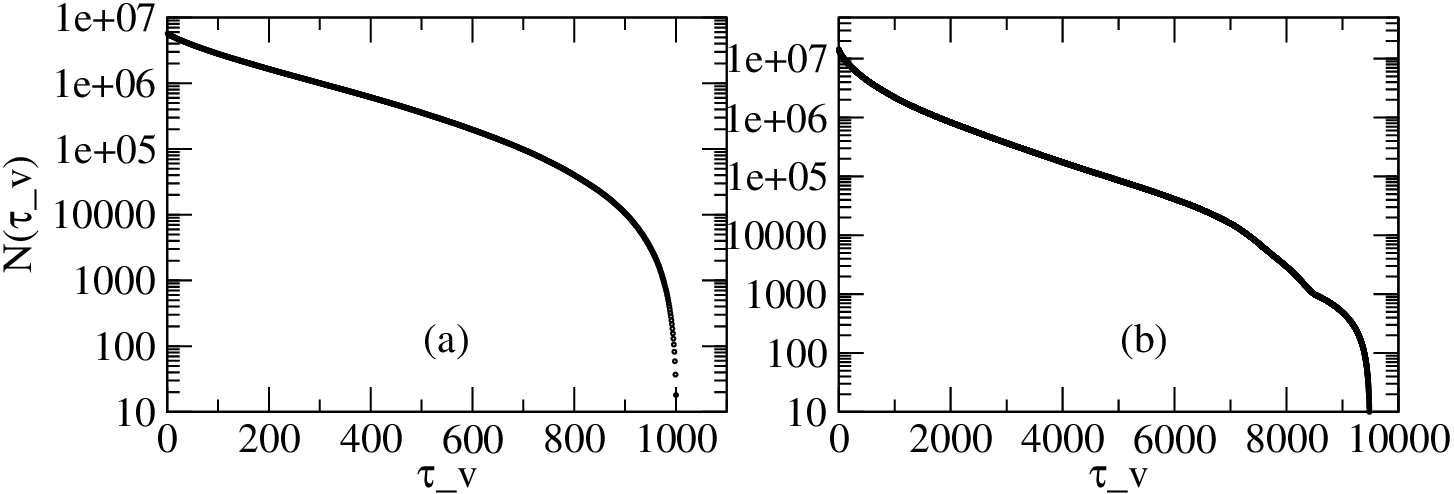} 
\caption{The distribution $N(\tau_v)$ for: (a) PKN Orlen (WSE), (b) TPSA (WSE). \vspace{1cm}}
\label{wykres4bis}
\end{figure}

The distribution $N(\tau_v)$ was presented in fig. \ref{wykres4bis}. One can see that there is an inflection point within the range of the variable $v$. It resembles the shape of the function $N(l)$ from the fig. \ref{fg1} and \ref{fg2}, which also possesses an inflection point.

By analogy to the volume time (\ref{tau_v}) we can define the volatility time:
\begin{equation}
\tau_\sigma(t_i)=\tau_\sigma(t_{i-1})+\sigma(t_i),
\end{equation}
where $\sigma(t_i)$ is the local volatility at the time $t_i$.

\begin{figure}
\vspace{0.5cm}
\includegraphics[scale=0.5]{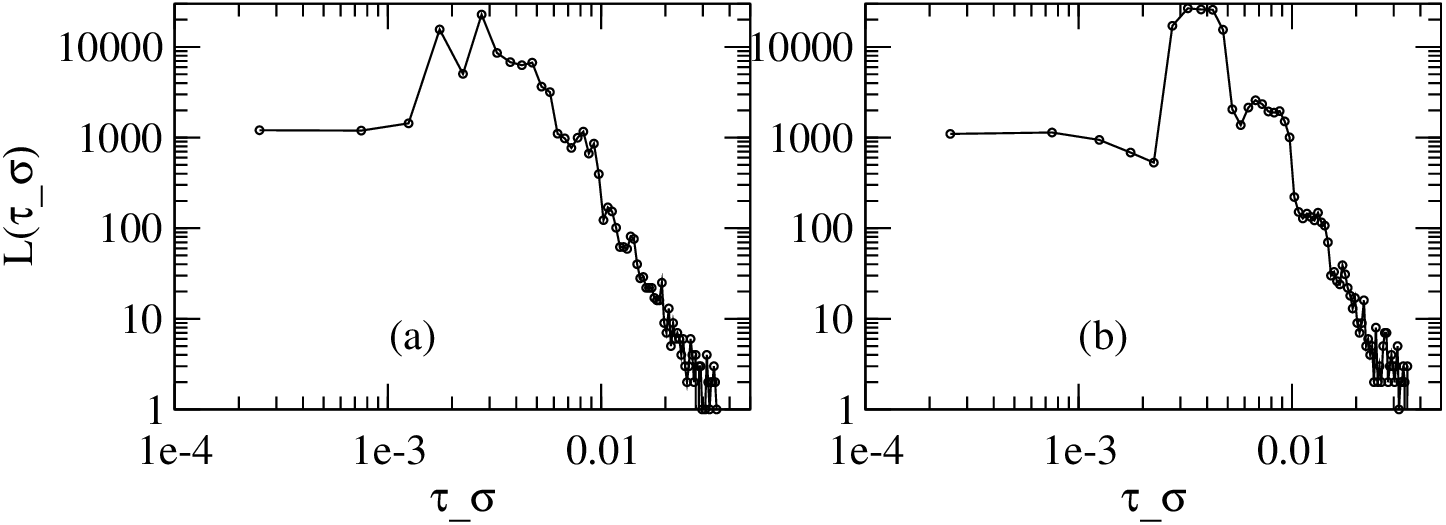} 
\caption{The binned distribution $L(\tau_\sigma)$, it is a number of trend periods of a length $\tau_\sigma$ measured in a volatility time, for: (a) PKN Orlen (WSE), (b) TPSA (WSE). \vspace{1cm}}
\label{wykres5}
\end{figure}

The volatility at time $t_i$ was defined as an absolute value of a log-return for a transaction at time $t_i$, it is
\begin{equation}
\sigma(t_i)=|\log(P(t_i)/P(t_{i-1}))|,
\end{equation} 
where $P(t_i)$ is the price of a stock at time $t_i$. For such a defined volatility we measured trend lengths and presented the distribution of the lengths $L(\sigma)$ in fig. \ref{wykres5}.

\section{Conclusions}

We have investigated short term price trends for high frequency stock market data. It turned out that the statistics for real markets is significantly different from the statistics of uncorrelated  processes. Longer trends (of the order of several minutes) are much more frequent than they should be, if one used an uncorrelated model. 

The investigations have been repeated for trends measured in volume and volatility time. The distribution of trends in volume time $N(\tau_v)$ has similar behaviour to the function $N(l)$.

We proposed a simple model that qualitatively captures the behaviour of the market.  The model leads to a distribution of trend series $N(l)$ that is similar to the distribution observed in market data. Our model produces also short range correlations. This behaviour is caused by the conditional probability of trend continuation that changes nonmonotonically with a trend length. At the beginning of the trend, the probability of the trend continuation grows, then it hits the maximum and finally  decreases. As a result, trends posses limited lengths.

\section*{Acknowledgments}
This work as a project of the COST Action P10 "Physics of Risk" was partially supported by Polish Ministry of Science and Higher Education, Grant No. 134/E-365/SPB/COST/KN/DWM 105/2005-2007.

\end{document}